\documentclass[aps,prl,twocolumn,superscriptaddress,longbibliography,showkeys]{revtex4-1}
\usepackage[colorlinks=true,citecolor=blue]{hyperref} 
\usepackage{color}
\usepackage{dsfont}
\usepackage{amsmath}
\usepackage{amssymb}
\usepackage{graphicx}
\usepackage{babel}

\usepackage{graphicx}
\usepackage{dcolumn}
\usepackage{bm}

\begin{document}

\title{Tensor network method for real-space topology in quasicrystal Chern mosaics}

\author{Tiago V. C. Ant\~ao}
\affiliation{Department of Applied Physics, Aalto University, 02150 Espoo, Finland}

\author{Yitao Sun}
\affiliation{Department of Applied Physics, Aalto University, 02150 Espoo, Finland}

\author{Adolfo O. Fumega}
\affiliation{Department of Applied Physics, Aalto University, 02150 Espoo, Finland}

\author{Jose L. Lado}
\affiliation{Department of Applied Physics, Aalto University, 02150 Espoo, Finland}

\date{\today}

\begin{abstract}
Computing topological invariants in two-dimensional quasicrystals and super-moire matter is a remarkable
open challenge,
due to the absence of translational symmetry and the colossal number of sites inherent to these systems.
Here, we establish a method to compute local topological invariants of exceptionally large systems using tensor networks, 
enabling the computation of invariants for Hamiltonians with hundreds of millions of sites, several orders of magnitude above
the capabilities of conventional methodologies. 
Our approach leverages a tensor-network representation of the density matrix using a Chebyshev tensor network algorithm, enabling large-scale calculations of topological markers in quasicrystalline and moire systems.
We demonstrate our methodology with two-dimensional quasicrystals featuring $C_8$ and $C_{10}$ rotational symmetries and
mosaics of Chern phases. 
Our work establishes a powerful method to compute
topological phases in exceptionally large-scale topological systems,
providing the required tool to rationalize generic
supe-moire and quasicrystalline topological matter.
\end{abstract}

\maketitle

\textit{Introduction: } Topological phases, although typically characterized by global invariants like the Chern number, can emerge locally in systems lacking translational symmetry \cite{Bianco_2011, Kitaev_2006, Callum_2020, Sykes_2021, Sykes2022, Bau_2024, Chen_2023, Chen_2023b, Sahlberg_2023, Sourav_2024}, as well as due to cross-dimensional phenomena in quasicrystals \cite{Antao_2024, Miranda_2024}. 
Chern markers allow for characterizing so-called Chern mosaics: spatially alternating local Chern numbers in moire systems. These topological mosaics have recently been observed in twisted graphene multilayers, where local probes revealed micron-scale domains with alternating Chern numbers of +1 and -1  \cite{Grover_2022}. Topological features are also expected to arise in moire quasicrystals like 30$^\circ$ twisted bilayer graphene  \cite{Andrei_2021, Nuckolls_2024, Yu_2019, Xia_2025}, 
twisted graphene trilayers \cite{Uri2023,Hao2024}, twisted transition metal dichalcogenide bilayers \cite{Li2024, Tsang2024}, and
generic platforms exhibiting quasicrystallinity or criticality \cite{DomnguezCastro2019, Gonalves_2022, Goncalves_2023,Oliveira_2023,PhysRevResearch.1.033009,PhysRevResearch.3.013262}. Interfaces between regions with different local Chern markers are expected to host chiral edge modes, forming Chern networks that connect local topological characterization with emergent macroscopic functionalities \cite{Gilbert_2025}. However, computing local invariants in exceptionally
large systems is a remarkable challenge, as it requires full knowledge of the many-body density matrix, a highly dense object.

\begin{figure}[t!]
    \centering
    \includegraphics[width=\columnwidth]{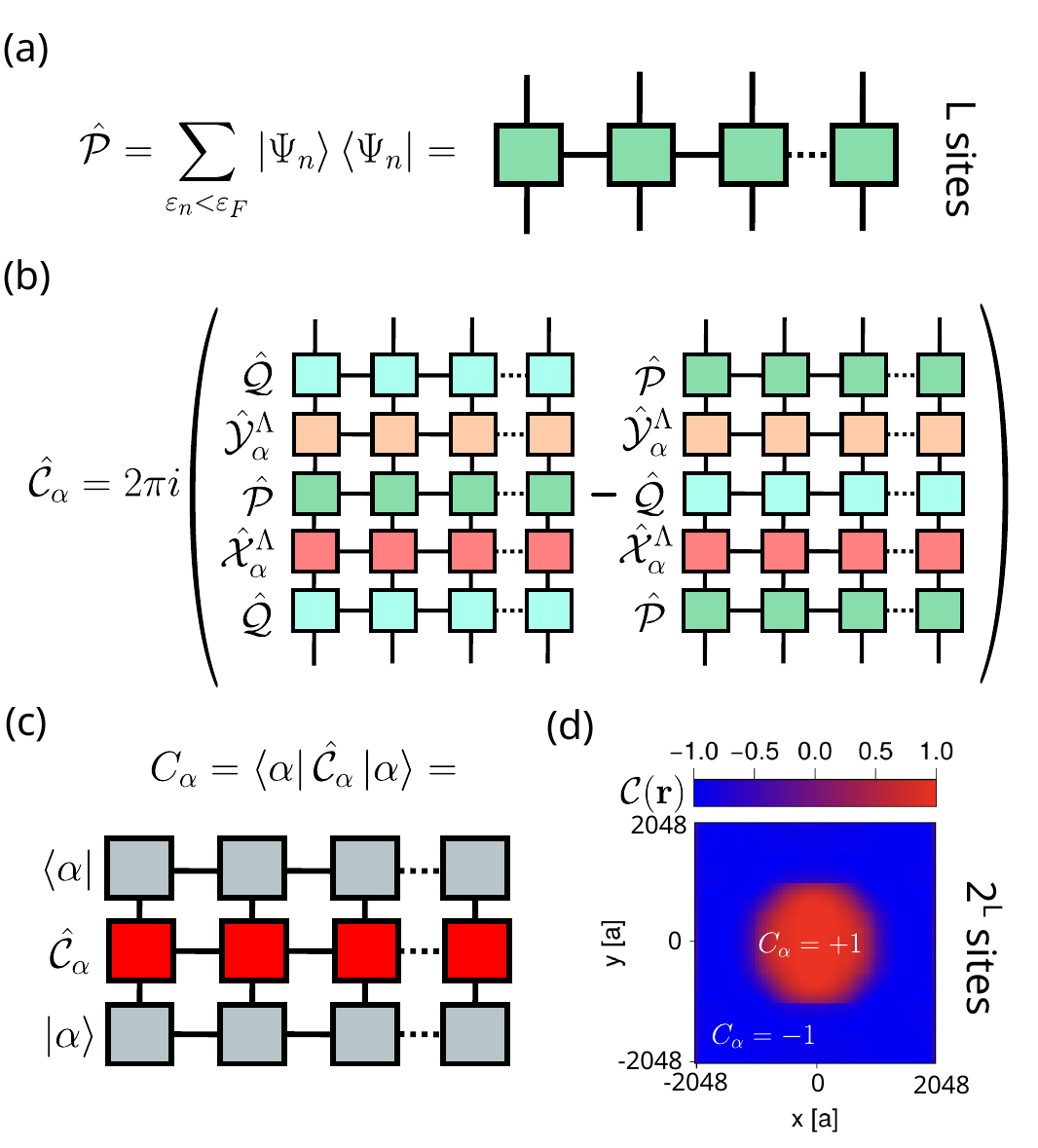}
    \caption{(a) Tensor network representation of the ground state projector,
    computed with a kernel polynomial method. (b) Tensor network construction of the local topological marker. (c) Extraction of the local Chern maker by tensor network contraction. (d) Demonstration for the topological marker for a modulated Chern system with different domains for a system with 16 million sites with a $C_\infty$ symmetric cosinusoidal modulation.}
    \label{fig:schematic}
\end{figure}

Quantum many-body Hamiltonians represent a paradigmatic problem featuring an exponentially large space,
whose exact solution becomes quickly unfeasible.
Tensor network representations of ground state wave-functions have enabled
obtaining exceptionally accurate solutions
to paradigmatic many-body problems \cite{ White_1992, Schollwck_2011, Schollwock_2005, Ors_2019, Paeckel_2019, Haegeman_2011, Niedermeier_2024, Lee_2016, PhysRevX.10.041038, Nakatani_2013, Szalay_2015, Chan2016, Huang2021, PhysRevLett.127.040507, PhysRevA.105.052446, fan2024overcomingzeroratehashingbound, Oseledets2010, Oseledets2011, Erpenbeck_2023, Murray_2024,PhysRevResearch.6.043182, PhysRevLett.130.100401, Holzner_2011},
providing a methodology to accurately solve many-body problems with an exponentially large Hilbert space.
Recently, tensor network techniques have been extended beyond the realm of quantum many-body physics,
finding applications in machine learning \cite{PhysRevResearch.4.043007,PhysRevX.8.031012,NIPS2016_5314b967},
complex dynamics \cite{Peddinti2024,Gourianov2025,niedermeierGP,boucomasGP,chenGP,2025arXiv250701149C},
ultra-precise function integration \cite{fernández2024learningtensornetworkstensor,PhysRevX.12.041018,PhysRevB.110.035124,PhysRevLett.132.056501},
and exceptionally large electronic problems \cite{Fumega2024,Sun_2025}.
The ability of tensor networks to represent functions on ultra-fine grids,
makes them a promising technique to compute observables that
require storing dense objects in exceptionally large systems.

Here, we establish a tensor network methodology enabling the computation of local topological markers in electronic models of unprecedented size, and
demonstrate the method for quasicrystalline Chern insulators.
This strategy relies on mapping the real space tight-binding electronic model
to a many-body pseudospin model that is solved with a tensor network kernel polynomial method.
This formalism allows for the computation of topological markers in systems with a Hamiltonian
too large to be explicitly stored,
reaching system sizes on the scale of microns. 
We demonstrate our methodology  depicted in Fig. \ref{fig:schematic} by applying it to real-space $C_8$ and $C_{10}$ symmetric modulated $\pi$-flux quasicrystal lattices. These modulations result in Chern mosaics emerging across the entire real space lattice. 
Our work overcomes the finite-size barrier of previous studies for the calculation of local topological invariants, as we perform calculations in systems with up to 268 million sites. We thus introduce a broadly applicable tool for exploring local topological order in quasicrystals and other moire-patterned quantum materials.

\textit{Tensor netowrk representation of the Chern marker:}
Our objective is to compute a real space topological marker for a large tight-binding Hamiltonian of the form
$
H = \sum_{\alpha \beta} t_{\alpha \beta} c^\dagger_\alpha c_\beta 
$
where $c^\dagger_i, c_i$ are the creation and annihilation operators in site $i$. In general, we will focus
on Hamiltonians $H$ that do not have translational symmetry, such as those arising from
a quasicrystalline modulation of the hoppings due to incomensurate super-moire patterns.
For systems with hundreds of millions of sites, $H$
becomes too large to even be stored. Using our methodology, we reinterpret a $2^L$-site
Hamiltonian as a many-body operator of an auxiliary many-body pseudo-spin chain of $L$ sites,
for which the index $s,s'$ corresponds to an element of a pseudo-spin many-body basis 
as $ s \equiv (s_1,s_2,..,s_L)$
and
$ s' \equiv (s'_1,s'_2,..,s'_L)$. In this form, the Hamiltonian
can be written using a tensor-network representation as a matrix product operator (MPO) in the pseudo-spin basis as \cite{Sun_2025}
\begin{equation}
\hat{\mathcal{H}} \equiv \sum \Gamma_{s_1,s_1'}^{(1)} \Gamma_{s_2,s_2'}^{(2)} \Gamma_{s_3,s_3'}^{(3)}\cdots\Gamma^{(L)}_{s_L,s_L'} c^\dagger_s c_{s'},
  \end{equation}
where $\Gamma_{s_i,s_i'}^{(n)}$ are tensors of dimension $\chi_n$, the bond dimensions of the MPO  \cite{itensor}.


Topological invariants, typically defined in momentum space through quantities like the Chern number \cite{PhysRevLett.49.405,RevModPhys.82.3045}, fail in non-periodic systems such as quasicrystals or disordered lattices, where real-space markers instead capture spatially resolved topological order \cite{Bianco_2011,Kitaev_2006,PhysRevLett.123.246801,2025arXiv250718919C}, remaining applicable to disordered, interacting, and higher-dimensional systems \cite{Chen_2023}, as well as real-space formulations of quantum geometry \cite{Peotta2015,Trm2022,PhysRevB.111.134201,PhysRevB.107.205133, Oliveira_2024}. 

The usual Chern marker is defined at any position $\alpha$ at the cost of the local spectral projectors into occupied states $\hat{P}$ and unoccupied states $\hat{Q}=\hat{I}-\hat{P}$ as

\begin{equation} C_\alpha = 2\pi i \langle \alpha |\hat{Q}\hat{X}\hat{P}\hat{Y}\hat{Q}-\hat{P}\hat{X}\hat{Q}\hat{Y}\hat{P}|\alpha \rangle, 
\label{eq:Chern}
\end{equation}
where $\hat{X}$ and $\hat{Y}$ are the position operators. In translationally invariant systems, $C_\alpha$ converges to the global Chern number in the bulk, while for inhomogeneous systems, $C_\alpha$ reveals the spatial structure of topological order, including the presence of different topological domains, leading to the possibility of diagnosing the presence of a Chern mosaic.

In the following, we cast this local Chern marker into a tensor network representation by evaluating the ground-state projector $\hat{\mathcal{P}}$ via a 
Chebyshev expansion using tensor networks \cite{PhysRevB.83.195115,PhysRevResearch.1.033009,PhysRevLett.130.100401}.
At zero temperature, the ground state projector is in exact correspondence with the density-matrix, which can be written as an MPO

\begin{equation}
    \hat{\mathcal{P}}=\int_{\varepsilon_F}^{\infty}  \delta(\omega-\hat{\mathcal{H}}) d\omega
    = \sum \Xi_{s_1,s_1'}^{(1)} \Xi_{s_2,s_2'}^{(2)}\cdots\Xi^{(L)}_{s_L,s_L'} |s \rangle \langle s' |,
\end{equation}
where $\Xi_{s_1,s_1'}^{(n)}$ are the tensors 
representing 
$\hat{\mathcal{P}}$, as graphically represented in Fig. \ref{fig:schematic}(a).
The explicit construction of the ground state projector
with usual matrices is highly demanding. For sparse Hamiltonians, Chebyshev methods allow for calculation of the local density of states (LDoS) at specific sites, since it can be computed using Chebyshev moments $\mu_n^\alpha$ locally at a site $\alpha$ using matrix-vector contractions. However, storing the full projector $\hat{P}$ is impossible even for modestly sized systems since the operator $\hat{P}$ is dense and therefore takes up memory which scales quadratically in the total number of atoms of the system. For example, for a  lattice of $10^3\times10^3$ atoms, one would generically need to store one trillion complex numbers and do algebra operations with a dense matrix of size $10^6\times10^6$. Using our method, however, the MPO representation is achievable both for sparse and dense matrices, and hence, one can make use of it to create a compressed and computationally manageable version of the density matrix $\hat{\mathcal{P}}$ which can be stored as a tensor network taking up only logarithmically increasing memory as a function of system size. Here, we  focus on $\pi$-flux quasicrystalline square lattice systems with up to 268 million sites.

We now elaborate on the construction of the Hamiltonian as a tensor network.
A generic single-particle tight-binding Hamiltonian can be written in tensor-network form as
$
    \hat{\mathcal{H}} = \sum_n\hat{\mathcal{B}}_n\mathbb{A}(n) + h.c., 
$
where $\mathbb{A}(n)$ corresponds to an MPO representation of the matrix with elements $\delta_{i,i+n}$, and $\hat{\mathcal{B}_n}$ corresponds to a diagonal MPO representing a matrix with elements $\left< i\right|\hat{\mathcal{B}}_n\left|i\right>$=$\hat{H}_{i,i+n}$. For instance, in the simple case where the Hamiltonian corresponds to a one-dimensional tight-binding chain, $\hat{\mathcal{B}}_n$ would yield the $n$th nearest neighbour hoppings. In two or higher-dimensional models, it becomes necessary to reshape the lattice into a periodically broken 1D chain with long range hoppings. The structure of the off-diagonals of the Hamiltonian matrix must therefore be reflected in the MPO $\hat{\mathcal{B}}_n$. In general, these MPOs can be constructed in one of two ways: (i) Constant or periodically broken MPOs can be constructed exactly as outlined below; (ii) Modulated or otherwise spatially varying structures can be constructed by mapping the real-space structure of the desired Hamiltonian off-diagonal and constructing a function $B_n(i)=H_{i,i+n}$, from which the MPO can be recovered using the QTCI algorithm \cite{Fernandez_2022, Ritter_2024, Fernandez_2024, Fernandez_2024b}. In essence, this approach combines a binary encoding of lattice coordinates with low-rank tensor interpolation, yielding an exponentially large effective grid at only polynomial cost.  Furthermore, the $\mathbb{A}(n)$ operators can be constructed for any $n$: The $1$-site translation operator can be defined explicitly using $\mathbb{A}(1)=\sum_{l}^L\left(\sigma_{l}^+ \bigotimes_{m > l} \sigma_{m}^-\right)$. From this expression, all $\mathbb{A}(n)$ operators can be calculated in polynomial time by decomposing $n = \sum_i 2^{l_i}$, and computing $\mathbb{A}\left(2^{l_i}\right)$ recursively as $\mathbb{A}\left(2^{l_i}\right)=\mathbb{A}\left(2^{l_i-1}\right)\cdot\mathbb{A}\left(2^{l_i-1}\right)$, which in the worst case scenario has to be done $L-1$ times. After this recursive calculation, one can simply take the product $\mathbb{A}(n)=\prod_i \mathbb{A}(2^{l_i})$.  In the case of a two-dimensional model, for instance, one can think of a $2^{L_x}\times2^{L_y}$ Hamiltonian as a chain, periodically interrupted every $2^{L_x}$ sites, with long range neighbours connecting over a distance of $2^{L_x}$ atoms. The Hamiltonian would contain two terms $\mathbb{A}(1)\mathbb{S}+\mathbb{A}(2^{L_x})$. The inclusion of the MPO $\mathbb{S}=\mathbb{I}-\bigotimes_l^L\mathbb{I}_l\prod_{m>L_y}^L \sigma_m^-$ produces the correctly interrupted hopping structure. Generically, 
this tensor-network representation allows us to perform the KPM expansion directly at the level of Chebyshev
polynomials of the Hamiltonian MPO $\hat{\mathcal{H}}$ \cite{PhysRevB.83.195115,PhysRevResearch.1.033009,PhysRevLett.130.100401,PhysRevResearch.6.043182,Fumega2024,Sun_2025} yielding the density matrix in MPO form as

\begin{equation}
    \hat{\mathcal{P}}=\sum_n T_n(\hat{\mathcal{H}})\int_{\varepsilon_F}^{\infty} d\omega \frac{T_n(\omega)}{\sqrt{1-\omega^2}},
\end{equation}
where $T_{n}(x)$ are the Chebyshev polynomials with
recursion relation $T_{n}(x) = 2xT_{n-1}(x) - T_{n-2}(x)$
with $T_0=1$ and $T_1 (x) =x$ \cite{RevModPhys.78.275}.

\begin{figure}[t!]
    \centering
    \includegraphics[width=\columnwidth]{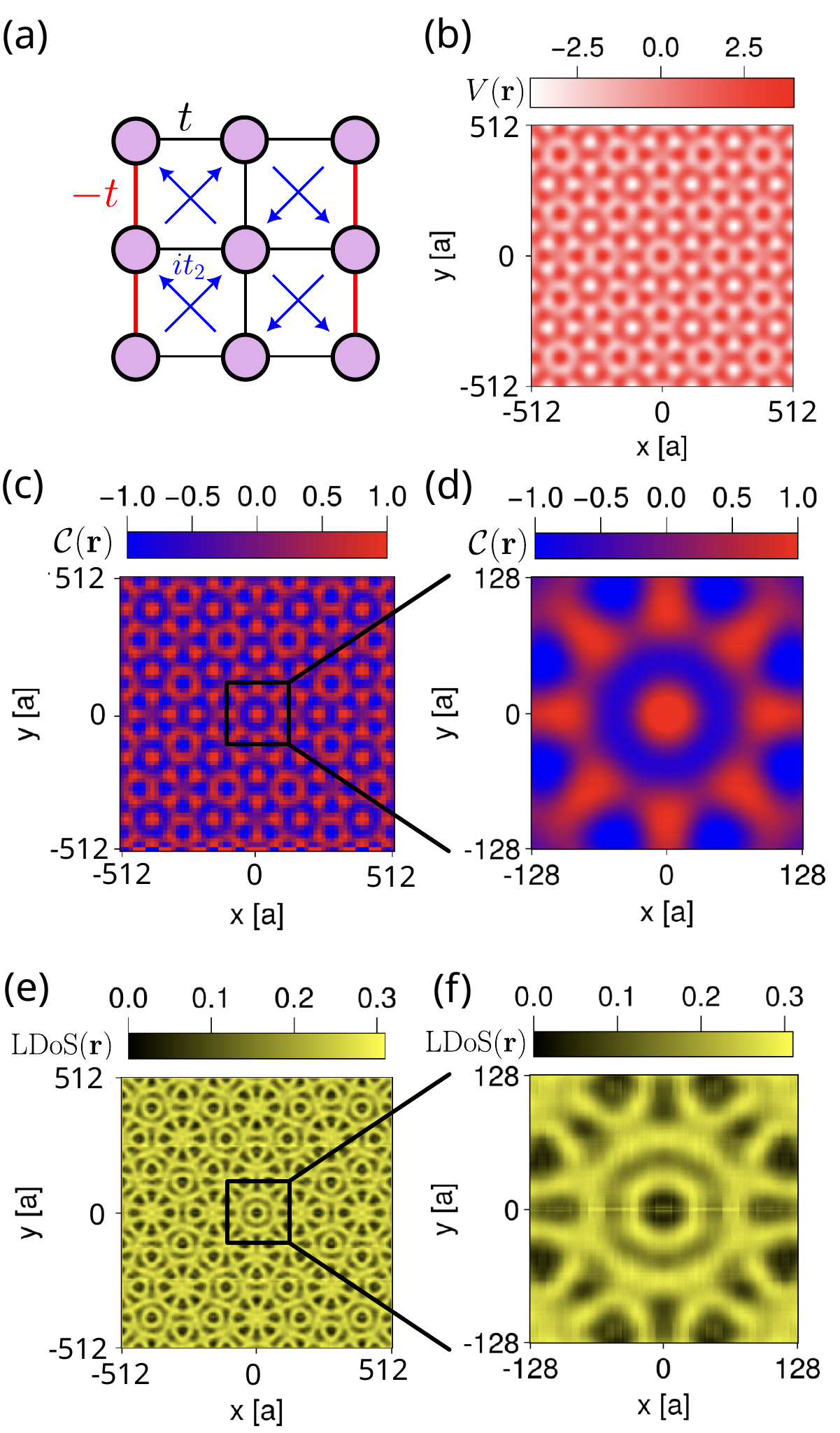}
    \caption{(a) Schematic representation of the $\pi-$flux model with nearest neighbour hopping $t$ and second nearest neighbour imaginary hopping $it_2$ (b) $C_8$ symmetric modulation of the second nearest neighbour hopping $t_2 = t_2^{(0)}V(\boldsymbol{r})$ (c) Quasicrystal Chern mosaic obtained from the MPO-KPM method across a $2^{10}\times2^{10}$ (1 million) atom lattice. (d) Zoom on the central $256\times 256$ atoms. (e) Local density of states of the Chern network resulting from the topological domains across the entire system. (f) Zoom on the central $256\times256$ atoms.}
    \label{fig:Ss_qs}
\end{figure}

We now elaborate on the tensor network representation of position operators.
We can make use of the QTCI algorithm \cite{QuanticsTCI.jl,TensorCrossInterpolation.jl} to  construct diagonal MPOs for the position operators $\hat{X}$ and $\hat{Y}$, yielding $\hat{\mathcal{X}}$ and $\hat{\mathcal{Y}}$ respectively. These could then be used to construct a topological marker MPO
$
    \hat{\mathcal{C}} = 2\pi i\left(
    \hat{\mathcal{Q}} \hat{\mathcal{X}} \hat{\mathcal{P}}
    \hat{\mathcal{Y}}
    \hat{\mathcal{Q}}-
    \hat{\mathcal{P}} \hat{\mathcal{X}} \hat{\mathcal{Q}}
    \hat{\mathcal{Y}}
    \hat{\mathcal{P}}\right)
$ in analogy with Eq. \ref{eq:Chern}.
However, while this expression can accurately be used with dense matrices and is mathematically exact, it is not optimal 
for tensor network calculations. Firstly, even though MPOs $\hat{\mathcal{X}}$ and $\hat{\mathcal{Y}}$ generically exhibit low bond dimensions when constructed from the QTCI algorithm, their products with the density matrix generate objects with a very large bond dimension, leading to difficulties with computational speed of calculations. On the other hand, for a tight-binding model with $2^L$ sites, the diagonal elements of $\hat{X}$ will themselves grow to $\mathcal{O}\left(2^{L-1}\right)$. This will lead to an amplification of errors 
stemming from the tensor network representing $\hat{\mathcal{P}}$,
by a factor that will increase when moving away from the center of the lattice, possibly accounting for the growth in bond dimension when computing $\hat{\mathcal{P}}$ or $\hat{\mathcal{Q}}$.
These issues can be resolved for tensor network based calculations by recasting the calculation of the global topological marker MPO with
quenched position operators. 
In particular, one can first center the position operators relative to the site $\alpha$ where one wishes to compute the marker's value. In addition one can bound the position operator by approximating $\hat{X}\approx \Lambda \sin(\hat{X}/\Lambda)$ for $x/\Lambda \ll 1$. The value of $\Lambda$ needs to be sufficiently large when compared to the correlation length of the topological operator such that the small angle approximation holds for a region which is wide enough to capture the local environment of the site $i$ correctly when taking products involving $\hat{\mathcal{P}}$ and $\hat{\mathcal{Q}}$ \footnote{We take a value of $\Lambda = 10a$, which leads to converged Chern markers.}.
This supresses the increase in bond dimension while retaining enough information to accurately predict the Chern marker.
Replacing $\hat{\mathcal{X}}\to\widetilde{\mathcal{X}}_\alpha^\Lambda=\Lambda\sin\left(\left(\hat{\mathcal{X}}-x_\alpha\right)/\Lambda\right)$ and equivalently $\hat{\mathcal{Y}}\to\widetilde{\mathcal{Y}}_\alpha^\Lambda$, we can approximately compute the local Chern marker as
\begin{equation}
    C_\alpha = 2\pi i\left<\alpha\right|
    \hat{\mathcal{Q}} \widetilde{\mathcal{X}}_\alpha^\Lambda \hat{\mathcal{P}}
    \widetilde{\mathcal{Y}}_\alpha^\Lambda
    \hat{\mathcal{Q}}-
    \hat{\mathcal{P}} \widetilde{\mathcal{X}}_\alpha^\Lambda\hat{\mathcal{Q}}
    \widetilde{\mathcal{Y}}_\alpha^\Lambda
    \hat{\mathcal{P}}\left|\alpha\right>,
\end{equation}
or as graphically depicted in Fig. \ref{fig:schematic}(b, c) and therefore resolve $C_\alpha$ across systems with hundreds of millions
of sites. An example for a simple circular sinusoidal modulation is given in Fig. \ref{fig:schematic}(d).

\begin{figure}[t!]
    \centering
    \includegraphics[width=\columnwidth]{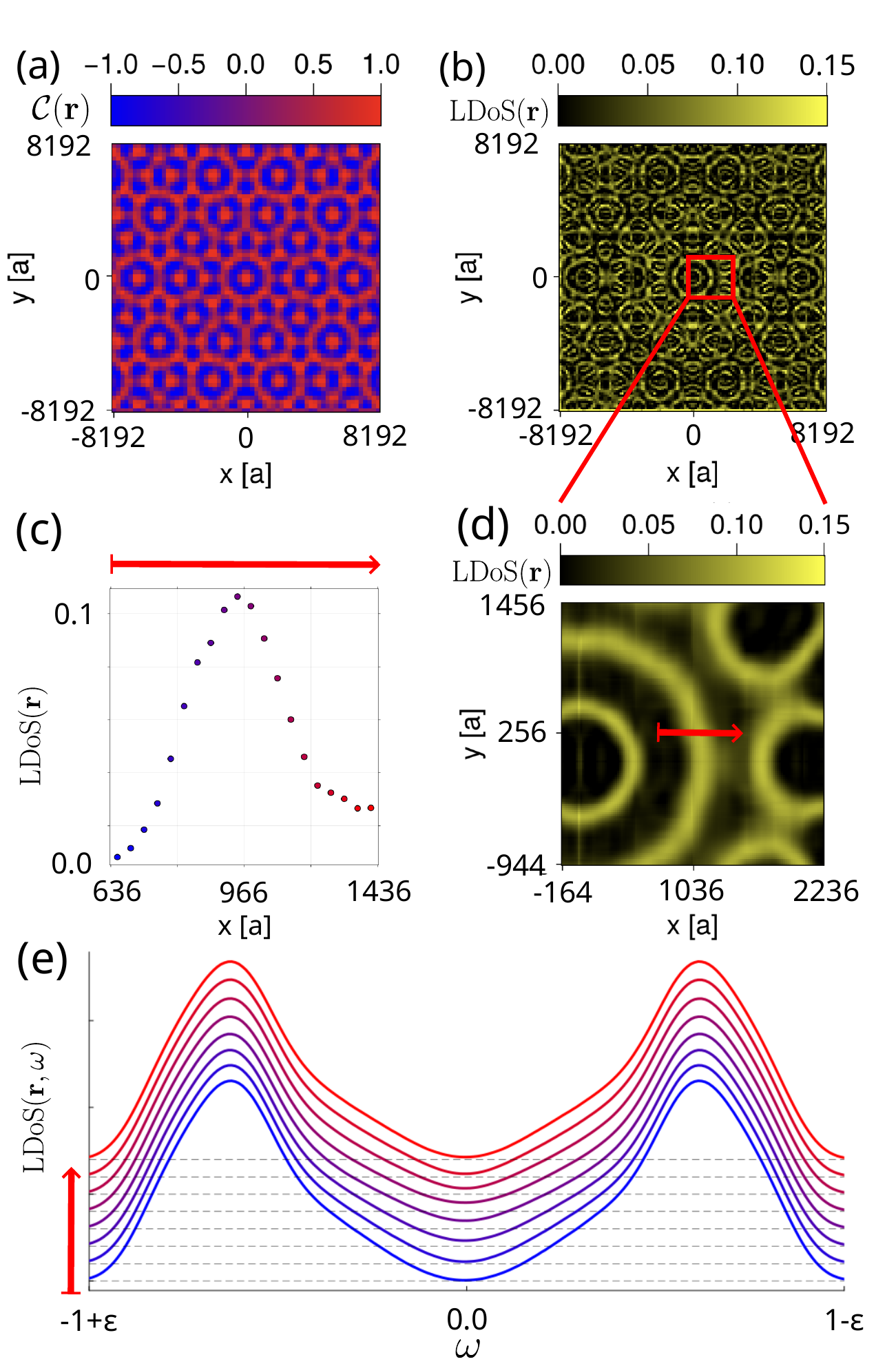}
    \caption{(a) Local Chern Marker obtained from the MPO-KPM method across a $C_{10}$ symmetric modulated lattice of $2^{14}\times2^{14}$ (268 million) atoms. (b) In-gap LDoS showcasing the presence of Chern network with a $C_{10}$ symmetric pattern. (c) LDoS along a path crossing an edge state. (d) Zoom in on a 2400$\times$2400 atom region containing the path of panel (c). (e) LDoS as a function of frequency for points along the same path calculated using KPM \cite{Sun_2025}}
    \label{fig:Chern_network}
\end{figure}

\textit{Quasicrystalline Chern Mosaic:}
We now use the previous methodology to resolve the topological domains of an exceptionally large
two-dimensional quasicrystal.
The $\pi$-flux model on the square lattice provides a paradigmatic realization of Dirac fermions in a two-dimensional system \cite{Goldman_2013}. In this model, each plaquette of the square lattice is threaded by a magnetic flux of $\pi$, 
leading to a nontrivial hopping phase structure, whose Hamiltonian reads
$H_1 = -t \sum_{\langle \alpha,\beta \rangle} e^{i A_{\alpha\beta}} c_\alpha^\dagger c_\beta + \text{h.c.}$. 
A convenient choice of gauge $A_{\alpha\beta}$  assigns alternating signs to the hoppings along one direction
as shown in Fig. \ref{fig:Ss_qs}(a). For a periodic system, thisT yields a Dirac model with two cones, for which topological phases can be induced by adding next-nearest-neighbor imaginary hoppings that break time-reversal symmetry.
as $H_2 = i t_2 \sum_{\langle\langle \alpha,\beta \rangle\rangle} \nu_{\alpha\beta} c_\alpha^\dagger c_\beta, $
where $\nu_{\alpha\beta} = \pm 1$ depends on the hopping direction. This term opens a gap at the Dirac points and turns the system into a Chern insulator
with Chern number $C = \text{sign}(t_2)$. 
We now apply our method to compute the local Chern marker to modulated $\pi-$flux models.  
We take $t_2 = t_2^{(0)}V(\mathbf{r})$ with 
$
    V(\mathbf{r})=V_0\sum_{i=1}^n \cos(\mathbf{q}_i\cdot\boldsymbol{r}),
$
with $t_2^{(0)}=0.2t$ and $V_0=1/2$, and set $n=4$ or $5$, corresponding to $C_8$ and $C_{10}$ symmetries. We use a modulation period of about 80 atomic sites for Fig. \ref{fig:Ss_qs}, such that $|\mathbf{q}_i|=2\pi/80$, yielding topological domains with characeristic width on the order of 40 atoms. For Fig. \ref{fig:Chern_network} we use a modulation period of $1080$ atoms. In both Fig. \ref{fig:Ss_qs}(c,d) for the $C_8$ modulation and Fig. \ref{fig:Chern_network}(a) for the $C_{10}$ case, we observe a quantization of the local marker to a very good degree. In general, the results reveal the presence of quasicrystalline Chern mosaics emerging due to the incommensurate geometry.

\textit{Quasicrystalline Chern Network:}
We now address the topologically protected edge states
occurring at every boundary between topological domains
in the quasicrystal Chern mosaic. At half-filling within the current model, the intricate nature of the $C_8$ and $C_{10}$ symmetric quasicrystalline Chern networks become evident in the local density of states, shown in Fig. \ref{fig:Ss_qs}(e,f) and Fig. \ref{fig:Chern_network}(b,c,d). This Chern network results in an enhancement of the in-gap density of states due to the presence of chiral in-gap edge states propagating along the bulk of the sample, as seen in Fig. \ref{fig:Chern_network}(c,e). Furthermore, it is worth noting that this Chern network percolates through the entire sample in Fig. \ref{fig:Ss_qs}(e,f) and Fig. \ref{fig:Chern_network}(b,d),
which would enable topologically protected conduction channels at a macroscopic scale. 
 
\textit{Conclusion:}
Computing topological invariants of non-periodic quantum systems,
including quasicrystal and
super-moire materials is an exceptional challenge due to the large sizes of these systems.
Here we have demonstrated a tensor network methodology that
enables computing local Chern number in real space for exceptionally
large models. Our methodology
relies on creating a tensor network representation of the topological
marker with a tensor network
Chebyshev expansion of the Hamiltonian. This methodology enables
computing topological invariants of exceptionally large models, and
in particular models whose single particle Hamiltonian
is too large to be stored explicitly.
We exemplified our methodology with $C_8$ and $C_{10}$ symmetric quasicrystals,
demonstrating that our methodology  enables faithfully
computing topological domains, whose impact
can be directly observed in the resulting network of topological
in-gap modes. Our approach can be extended to compute topological invariants in super-moire systems, including valley Chern numbers, $\mathbb{Z}_2$ quantum spin Hall indices, crystalline Chern numbers, and real-space quantum geometry, providing a general framework for analyzing topological states in large-scale super-moire matter.


\textbf{Acknowledgments:}
We acknowledge the computational resources provided by the Aalto Science-IT project
and the financial support from the Academy of Finland Projects Nos. 331342, 358088, and 349696,
the Jane and Aatos Erkko Foundation, the Finnish Quantum Flagship,  
InstituteQ and ERC Consolidator Grant ULTRATWISTROICS (Grant agreement no. 101170477). 
We thank X. Waintal, M. Niedermeier, 
I. Sahlberg, R. Oliveira, M. Nguyen,
T. Heikkil\"{a}, P. T\"{o}rm\"{a} and T. Ojanen
for useful discussions.

\bibliography{biblio}

\end{document}